\documentclass[fleqn,usenatbib]{mnras}

\usepackage{newtxtext,newtxmath}

\usepackage[T1]{fontenc}

\usepackage{mathrsfs}
\usepackage{ulem}

\usepackage{graphicx}	
\usepackage{amsmath}	

\usepackage{hyperref}
\title[Properties of the bars by the bar instability]
{Intrinsic properties of the bars formed by the bar instability in
flat stellar discs}

\author[Shunsuke Hozumi]{
Shunsuke Hozumi\thanks{E-mail: hozumi@edu.shiga-u.ac.jp}\\
Faculty of Education, Shiga University, 2-5-1 Hiratsu,
Otsu, Shiga 520-0862, Japan
}

\date{Accepted 2021 December 13.
      Received 2021 December 1; in original form 2021 July 24}

\pubyear{2021}

\begin{document}
\label{firstpage}
\pagerange{\pageref{firstpage}--\pageref{lastpage}}
\maketitle

\begin{abstract}
The properties of the bars formed by the bar instability
are examined for flat stellar discs. The initial mass models
chosen are Kuzmin--Toomre discs, for which two types
of exact equilibrium distribution function (DF) are
employed in order to realize different distributions
of Toomre's $Q$ values along the radius. First, the most
linearly unstable, global two-armed modes (MLUGTAMs) of
these disc models are determined by numerically solving
the linearized collisionless Boltzmann equation. Next,
we carry out $N$-body simulations whose models are
constructed from the DFs adopted above. The latter
simulations unravel that the MLUGTAMs corresponding
to those obtained from the former modal calculations
are excited in the early phases of evolution, finally
being deformed into bars in the nonlinear regime by
the bar instability. We show that for simulated bars,
the length increases and the axis ratio, in essence,
decreases as the amplitude increases. These correlations
are almost similar to those of the observed bars. In addition,
we find that these bar properties are tightly correlated
with the initial typical $Q$ value, irrespective of the DF.
In conclusion, a disc with a smaller typical $Q$ value produces
a bar which is smaller in amplitude, shorter in length and
rounder in shape. This finding might suggest that the Hubble
sequence for barred galaxies is the sequence of decreasing
$Q$ from SBa to SBc or SBd. The implied correlations between
the initial typical $Q$ value and each of the bar properties
are discussed on the basis of the characteristics of the MLUGTAMs.
\end{abstract}




\section{Introduction}\label{sec:introduction}
Spiral and barred structures are the outstanding features in
disc galaxies. Above all, bars can play an important role in
the central activities and the secular evolution of disc galaxies
as well as the dynamical one. For example, a bar is regarded
as a key agent effective in extracting angular momentum from
the gas in a disc, and as a result, also effective in fuelling
the gas into the nucleus \citep{lia92, wh95, panos00, jsk05,
sheth05}. Eventually, such gas inflow could drive the central
activities of disc galaxies \citep{fbk94, ellison11}. In addition,
a bar as a wave pattern can cause wave-particle interactions with
halo particles, so that the disc angular momentum is transferred
to the halo \citep{lia02, lia03, victor06, sellwood14}. In this
manner, the existence of a bar propels the secular evolution
of a disc through such an angular momentum redistribution.
The details of the phenomena mentioned above will hinge
on the bar properties. In fact, as demonstrated from numerical
simulations, the efficiency of transferring the gas to the
centre of a galaxy along its bar depends sensitively on the
bar strength that is one of the significant bar properties
\citep{lia92, fb93, rt04}. Accordingly, elucidating the
origin of the bar properties will finally lead to a better
understanding of the complete picture of how barred galaxies
evolve inherently.

From an observational point of view, the fraction of barred
galaxies amounts to more than two-thirds of the entire disc
galaxies, if not only strongly but also weakly barred galaxies
are included \citep{ksp2000, eskridge00, lsb04, menendez07,
mj07, barazza08, amc09, buta15, diaz16a, diaz16b}.
As a note, \citet{block91} found that the bar fraction is more
enhanced in near infrared than in optical wavelengths, and subsequent
related studies reinforce their finding \citep{grosbol98,marquez99,
eskridge00,menendez07,lsbk11}. In addition,
\citet{sheth08} and \citet{melvin14} have revealed that the
bar fraction increases with time over the last 7 or 8 Gyr.
On the other hand, \citet{elmegreen04}
and \citet{jogee04} advocate that the observed bar fraction
is nearly constant over time, and \citet{simmons14}
have also reported that the bar fraction in
the selected redshift range of $0.5\le z\le 2$ shows no significant
evolution. In either case, barred galaxies outnumber non-barred galaxies.
This fact may indicate that galactic discs are prone to form
bars, the reason of which might be found out by investigating
how the bar properties are acquired.

On the background of an increasing number of observed barred
galaxies, some properties of galactic bars have been uncovered
on a statistically meaningful level of confidence. For instance,
\citet{erwin05} has shown that bars in S0-Sb galaxies are longer
than those in Sc-Sd galaxies, which was pointed out before
with relatively small samples in the literature \citep[e.g.][]
{martin95, ee85, chapelon99, laine02, ls02, lsr02}. Subsequently,
\citet{eekb07} noticed a positive correlation between the
length and the amplitude of bars. Thereafter, similar results
have been presented by \citet{diaz16a}, \citet{guo19} and
\citet{cuomo19, cuomo20}. Moreover, \citet{menendez07} and
\citet{hoyle11} unravelled that larger amplitude bars are
more elongated in shape. These correlations
might be formed by the difference in evolving speed. That is,
because galaxies with higher central densities transfer angular
momentum faster from a bar to the surrounding components such
as a disc and a halo, a bar changes its properties over time
\citep{eekb07, menendez07, hoyle11}. However, this line of
explanation does not necessarily exclude the view that
the bar properties are different at the epoch of bar
formation owing to the difference in the physical properties
of the galactic discs that lead to barred structures in the end.


As mentioned above, recent observations have exposed the
detailed properties of bars for each class of disc galaxies
along the Hubble sequence, while the mechanism of bar
formation itself is not yet fully understood. Empirically,
we know from numerical simulations that galactic discs
are dynamically unstable to result in bars via the bar
instability \citep[e.g.][]{h71, sellwood81, lia84, sw93}.
On the other hand, as noted above, we also know that non-barred
galaxies do exist at no small fraction of disc galaxies
in the real Universe. Therefore, the central issue of
the disc dynamics in the past was to disclose how the discs
were stabilized to survive as unbarred. As a result, some
stability criteria of flat discs against bar formation were
proposed \citep{op73, eln82}, although they stood on an empirical
basis. In a reflection of this circumstance, we did not pay,
in essence, any attention to the bar properties generated by
the bar instability at that time. Recent $N$-body simulations
using reasonably realistic models with a huge number of particles
have made it possible to compare simulated bars with observed
ones, and they have revealed that the buckling instability
deforms a bar which is caused by the bar instability into
a boxy/peanut-shaped or X-shaped bulge. In fact, so-called
pseudobulges which exhibit boxy/peanut/X-shaped features have
been observed in real barred galaxies \citep{bureau99, lutticke00a,
lutticke00b, yy15, erwin17, panos21}. In particular, our Galaxy
and M31 are considered to have such pseudobulges \citep{dwek95,
lia06, gerhard12, portail17, ciambur17}. In this way, morphological
and kinematical similarities between simulated and observed bars
have been indicated in many studies \citep[e.g.][]{od03, panos06,
lia15, xiang21}. However, we have not sufficiently understood
the origin of the bar properties such as the correlation between
the length and the amplitude and that between
the axis ratio and the amplitude.


In relation to the destruction of a bar that is induced
by a massive central black hole, \citet{sh12} has revealed
that the bar properties arising from the bar instability
are well-correlated with a typical \citeauthor{toomre64}'s
(\citeyear{toomre64}) $Q$ value for a given functional form
of velocity distribution. That is, as a typical $Q$ value
is smaller, the resulting bar becomes lower in amplitude,
shorter in length, and rounder in shape. However, the origin
of these bar properties still remains unclear.

In this paper, we show that the properties of the bars are
closely connected to those of the most linearly unstable,
global two-armed modes (MLUGTAMs), on the basis of which
their origin is considered. In Section \ref{sec:models},
the disc models constructed from exact equilibrium
distribution functions (DFs) are described. In Section
\ref{sec:method}, we explain how to determine the MLUGTAM
and how to follow the evolution of an $N$-body model which
is realized from the DF. Results including correlations
between the MLUGTAMs and formed bars are presented in
Section \ref{sec:results}. We discuss the origin of the
bar properties in Section \ref{sec:discussion}. Conclusions
are given in Section \ref{sec:conclusions}.

\section{Models}\label{sec:models}
In order to highlight the bar properties arising from the bar
instability, we need to reduce likely factors which influence
the formation and evolution of bars. If approximate equilibrium
models were used, initial transients, for example, like those
revealed by \citet{fujii11}, could emerge in the early evolving
phases of numerical simulations, so that the subsequent disc
evolution and bar formation could be altered to some degree.
Therefore, the appropriate disc models should be constructed
on the basis of mathematically consistent and exact equilibrium
DFs. This reasoning forces us to handle two-dimensional
zero-thickness discs, although present-day computers have
enabled us to easily simulate three-dimensional disc galaxies
with a huge number of particles \citep[e.g.][]{dbs09, dvh13,
fujii18, fujii19}. This is because exact equilibrium DFs are
known only for infinitesimally thin discs.

We adopt razor-thin Kuzmin--Toomre (K--T) discs \citep{kuzmin56,
toomre63}, whose surface density, $\mu$, and potential, $\Phi$,
are, respectively, given by
\begin{equation}
\mu(r)=\frac{M_\mathrm{d}}{2\uppi a^2}
  \left(1+\frac{r^2}{a^2}\right)^{-3/2}
\label{eq:sd_disc}
\end{equation}
and
\begin{equation}
\Phi(r)=-\frac{GM_\mathrm{d}}{a}\left(1+\frac{r^2}{a^2}\right)^{-1/2},
\label{eq:pot_disc}
\end{equation}
where $M_\mathrm{d}$ and $a$ are, respectively, the mass and
scale length of the disc, $G$ is the gravitational constant,
and $r$ is the distance from the disc centre.

The equilibrium DFs of directly rotating (prograde) stars
for the K--T discs have been derived by \citet{miya71} and
by \citet{kalnajs76}. In general, for a flat axisymmetric
galaxy, the equilibrium DF of prograde stars, $F^+$, is
represented by $F^+(\varepsilon,\,j)$, where $\varepsilon$
and $j$ are, respectively, the energy and the angular
momentum of a star per unit mass. On the other hand,
there is no definite way of prescribing the distribution
of retrograde stars. Then, we introduce them with the
method devised by \citet{nishida84}, on the ground that
it includes no additional parameter. Consequently, the
equilibrium DF for both prograde and retrograde stars,
$F_0(\varepsilon,\,j)$, is written by
\begin{equation}
F_0(\varepsilon,j)=\left\{
  \begin{array}{ll}
    (1/2){F_0}^+(\varepsilon)+{F_1}^+(\varepsilon,\,j) & j\geq 0\\
    (1/2){F_0}^+(\varepsilon) & j<0,
  \end{array}\right.
\label{DF}
\end{equation}
where the functions ${F_0}^+(\varepsilon)$ and ${F_1}^+(\varepsilon,\,j)$
are derived from the expansion of $F^+(\varepsilon,\,j)$ with respect
to the angular momentum such that
\begin{equation}
F^+(\varepsilon,\,j)={F_0}^+(\varepsilon)+{F_1}^+(\varepsilon,\,j).
\end{equation}

\begin{figure}
\centerline{\includegraphics[width=6.5cm]{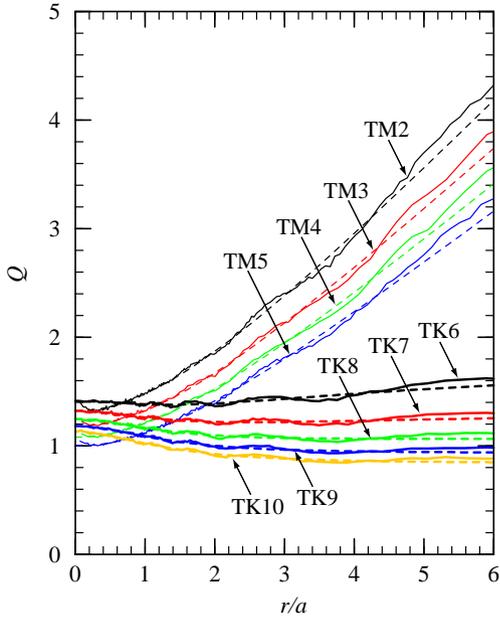}}
\caption{Toomre's $Q$ parameters as a function of radius which
is normalized by the scale length of the disc, $a$, for the
models with Kalnajs's DFs (thick lines) and those with Miyamoto's
DFs (thin lines). Solid lines show the $Q$ values calculated
from $N$-body realization of each model, while dashed lines
represent those calculated from corresponding analytical DFs.}
\label{fig:Q_profile}
\end{figure}

We construct $F^+(\varepsilon,\,j)$ by employing the approach
of \citet{miya71} and that of \citet{kalnajs76}. In both types
of DFs, there is a model parameter which prescribes the kinematic
structure of a disc. When denoting this parameter as $m_\mathrm{K}$
for Kalnajs's DFs and as $m_\mathrm{M}$ for Miyamoto's DFs, we take
$m_\mathrm{K}=6, 7, 8, 9,$ and 10, which are termed models TK6, TK7,
TK8, TK9, and TK10, respectively, and $m_\mathrm{M}=2, 3, 4,$ and
5, which are named models TM2, TM3, TM4, and TM5, respectively.
These DFs are used for determining the MLUGTAMs with no explicit
truncation of the discs, as described in Subsection
\ref{subsec:linear_mode}, while the initial phase-space
coordinates are realized with $N=10\,000\,000$ particles
of equal mass by truncating the discs at $r=10\,a$ in order
to examine the time evolution of these models with the method
explained in Subsection \ref{subsec:evolution}.

In Fig.~\ref{fig:Q_profile}, we present \citeauthor{toomre64}'s
(\citeyear{toomre64}) $Q$ profiles for all the above-mentioned
models realized with $N$ particles, along with those calculated
from the analytic forms of the corresponding DFs. The models
generated from Kalnajs's DFs have slightly declining $Q$
distributions with the radius except for model TK6 that
shows a slight increase in $Q$ from $r\sim 1.3\,a$ with
the radius, while those from Miyamoto's DFs show steeply
rising $Q$ distributions with the radius such that $Q\propto
\sqrt{(r/a)^2+4}$. At any rate, as the model parameters,
$m_\mathrm{M}$ and $m_\mathrm{K}$, increase, the $Q$ values
at all radii decrease for both DFs. In addition, as
found from Fig.~\ref{fig:Q_profile}, for all models but
models TK9 and TK10, the $Q$ values are larger than unity
throughout the radius, so that these discs are stable
against local axisymmetric Jeans instabilities at all
radii, while the $Q$ values for models TK9 and TK10
are lower than unity at large radii, so that these
two models are unstable at the corresponding radii.
In reality, regardless of the locally axisymmetric
stability, all models are bar-unstable as shown in
Subsection \ref{subsec:bars}.

\section{Method}\label{sec:method}
The origin of the bar properties formed by the bar instability
is considered to have a close connection to the MLUGTAMs,
since such modes are known to deform into bars in nonlinear
stages \citep[e.g.][]{nishida84}. Then, we first determine
the MLUGTAMs for the DFs of the models constructed in Section
\ref{sec:models} as a basis for the prescription of bars. Next,
the corresponding $N$-body models are evolved self-consistently
until a formed bar has relaxed to a steady state.

\subsection{Determination of the fastest growing two-armed modes}
\label{subsec:linear_mode}
The MLUGTAMs are determined
by numerically integrating the linearized collisionless Boltzmann
equation, which is represented by
\begin{equation}
\frac{\mathrm{d}f_m}{\mathrm{d}t}=\frac{\partial \psi_m}{\partial r}
\frac{\partial F_0}{\partial u}+\mathrm{i}\,m\psi_m
\frac{\partial F_0}{\partial j},
\label{eq:linear_cB}
\end{equation}
where the functions $f_m$ and $\psi_m$ are the $m$th Fourier
components of the perturbed part of the DF
and potential, respectively, $u$ is the radial velocity, and $t$
is the time.

Equation (\ref{eq:linear_cB}) is solved for $m=2$ modes as
an initial value problem. That is, we continue integrating
equation (\ref{eq:linear_cB}) with respect to the time until
an imposed perturbation has reached exponentially growing
phases in $f_m$. The details of the method for finding the
MLUGTAMs are described by \citet{hfn87} and by
\citet{hf89}.

\subsection{Evolution of stellar discs}\label{subsec:evolution}
The time evolution of the K--T discs is followed with a
self-consistent field (SCF) method \citep{ho92, hh95, sh97}.
As shown analytically by \citet{miller71, miller74} and
numerically by \citet{es95}, a softening length introduced
in conventional $N$-body techniques has a stabilizing effect
on stellar discs, so that, for example, the growth rate of
the MLUGTAM is reduced considerably as compared with that
obtained using unsoftened gravity, even though a relatively
small softening length is assigned. Since no such gravitational
softening is explicitly included in the SCF approach,
we can directly compare the growth rate and pattern
speed of the MLUGTAM estimated in an SCF simulation
with those determined by a softening-free phase-space
method of integrating equation (\ref{eq:linear_cB}).

For two-dimensional SCF simulations, we use the same method
as that adopted by \citet{hh05}, in which it is necessary to
prepare a pair of the density and potential basis functions,
$(\mu_{nm},\, \Phi_{nm})$, such that each pair satisfies
Poisson's equation given by
\begin{equation}
\nabla^2 \Phi_{nm}(\mathbfit{r})=4\uppi G\mu_{nm}(\mathbfit{r})\delta (z),
\end{equation}
where $\mathbfit{r}$ is the position vector in the disc
plane, $\delta$ is the delta function with $z$ being the vertical
coordinate to the disc plane, and $n$ and $m$ are
those positive integers or zero which indicate the order in the
radially and azimuthally expanded terms, respectively. In a set
of these basis functions, the density and potential of the system
can be expanded, respectively, in the forms
\begin{equation}
\mu(\mathbfit{r})=
\sum_{n=0}^{n_\mathrm{max}}\sum_{m=0}^{m_\mathrm{max}}
A_{nm}(t)\mu_{nm}(\mathbfit{r})
\end{equation}
and
\begin{equation}
\Phi(\mathbfit{r})=
\sum_{n=0}^{n_\mathrm{max}}\sum_{m=0}^{m_\mathrm{max}}
A_{nm}(t)\Phi_{nm}(\mathbfit{r}),
\end{equation}
where $A_{nm}(t)$ are the expansion coefficients at time $t$,
and $n_\mathrm{max}$ and $m_\mathrm{max}$ are the maximum numbers
of the radial and azimuthal expansion terms, respectively.
We operate $\Phi_{nm}$ to the particle distribution at every
time-step to obtain $A_{nm}(t)$ with the help of the bi-orthogonality
between $\mu_{nm}$ and $\Phi_{nm}$. Consequently, the accelerations,
$\mathbfit{a}(\mathbfit{r})$, are provided by
\begin{equation}
\mathbfit{a}(\mathbfit{r})
=-\nabla\Phi(\mathbfit{r})
=-\sum_{n=0}^{n_\mathrm{max}}\sum_{m=0}^{m_\mathrm{max}}
A_{nm}(t)\nabla\Phi_{nm}(\mathbfit{r}),
\end{equation}
where $\nabla\Phi_{nm}(\mathbfit{r})$ can be calculated
analytically in advance, when a basis set is specified. Once the
accelerations for all particles are evaluated, the equations of
motion are integrated in Cartesian coordinates with a time-centred
leap-frog algorithm \citep[e.g.][]{press86}.

The amplitude and pattern speed of the MLUGTAM at time $t$
are estimated in the same way as that carried out by \citet{hh05},
although the bar amplitude itself will be defined in Subsection
\ref{subsec:correlations}. Then, the amplitude is obtained from
the absolute value of the expansion coefficient, $|A_\mathrm{22}(t)|$,
and the pattern speed is calculated from half the time change
in the phase of $A_\mathrm{22}(t)$.

We adopt \citeauthor{ai78}'s (\citeyear{ai78}) basis set, which
is suitable for flat discs and is constructed on the basis of
the K--T discs, represented by
\begin{equation}
\mu_{nm}(\mathbfit{r})=\frac{M_\mathrm{d}}{2\uppi {a_\mathrm{b}}^2}
(2n+1)\left(\frac{1-\xi}{2}
\right)^{3/2}P_{nm}(\xi)\exp(\mathrm{i}\,m\varphi)
\label{eq:density_basis}
\end{equation}
and
\begin{equation}
\Phi_{nm}(\mathbfit{r})=-\frac{GM_\mathrm{d}}{a_\mathrm{b}}
\left(\frac{1-\xi}{2}\right)^{1/2}P_{nm}(\xi)\exp(\mathrm{i}\,m\varphi),
\label{eq:pot_basis}
\end{equation}
where $P_{nm}$ $(n\ge m)$ are the Legendre functions, $\varphi$ is the azimuthal
angle, and $\xi$ stands for the radial transformation defined as
\begin{equation}
\xi=\frac{r^2-{a_\mathrm{b}}^2}{r^2+{a_\mathrm{b}}^2}.
\label{eq:r_to_xi}
\end{equation}
In equations (\ref{eq:density_basis}), (\ref{eq:pot_basis}), and
(\ref{eq:r_to_xi}), the scale length of the basis set, $a_\mathrm{b}$,
is not necessarily required to be equal to that of the disc, $a$.
However, since we adopt the \mbox{K--T} discs whose functional
forms are identical to those of the lowest-order members of the
basis functions, we choose $a_\mathrm{b}=a$. For SCF simulations,
we set $n_\mathrm{max}=24$ and $m_\mathrm{max}=12$. In the angular
expansion, only even $m$-values are retained to extract the
properties of a bar easily. Furthermore, we have parallelized
the SCF code in accordance with the prescription given by
\citet{hsb95} \citep[see also][]{kelly01} to reduce computation
time.

We present results in the system of units such that
$\mbox{$G=1$},\, M_\mathrm{d}=1$, and $a=1$.
If unit time and velocity are converted into
those scaled to physical values appropriate for the Milky
Way, they become $9.77\times 10^6$ yr and 260 km$\,$s$^{-1}$,
respectively, using $M_{\rm d}=4.1\times 10^{10}$ M$_{\sun}$
and $a=2.6$ kpc \mbox{\citep{bg16}}. In this case, the unit
of angular speed is 100 km$\,$s$^{-1}\,$kpc$^{-1}$. These
values should be viewed as a reference, because the surface
density profile of the K--T discs is different from that of
real disc galaxies. The models are evolved forward in time
until $t=800$ at which the bar phases are sufficiently long
to be relaxed, with a time-step of 0.05 for all simulations.
As a result, the relative energy error was, in all cases,
smaller than 0.062 per cent.

\section{Results}\label{sec:results}
\subsection{Growth of linearly unstable two-armed modes}
\begin{figure}
\centerline{\includegraphics[width=7.5cm]{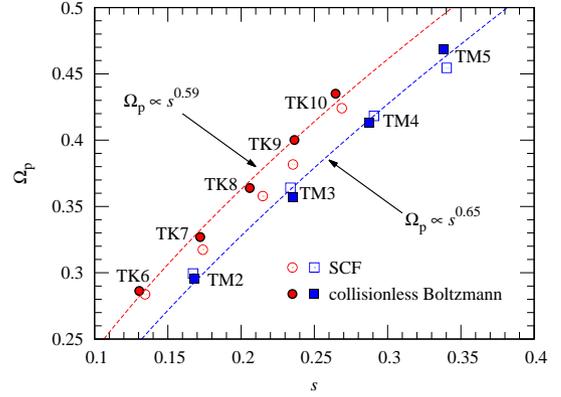}}
\caption{Growth rates, $s$, and pattern speeds,
$\Omega_\mathrm{p}$ of the most linearly unstable, global
two-armed modes for the models with Kalnajs's DFs (circles)
and those with Miyamoto's DFs (squares). Filled symbols
show the values obtained by solving the linearized collisionless
Boltzmann equation, while open symbols stand for those from the
SCF simulations. The dashed lines denote power-law fits for
the data obtained with the linear modal calculations. The
red dashed line represents the fit for Kalnajs's DFs while
the blue one expresses that for Miyamoto's DFs}
\label{fig:eigen_values}
\end{figure}

\begin{figure}
\centerline{\includegraphics[width=6.5cm]{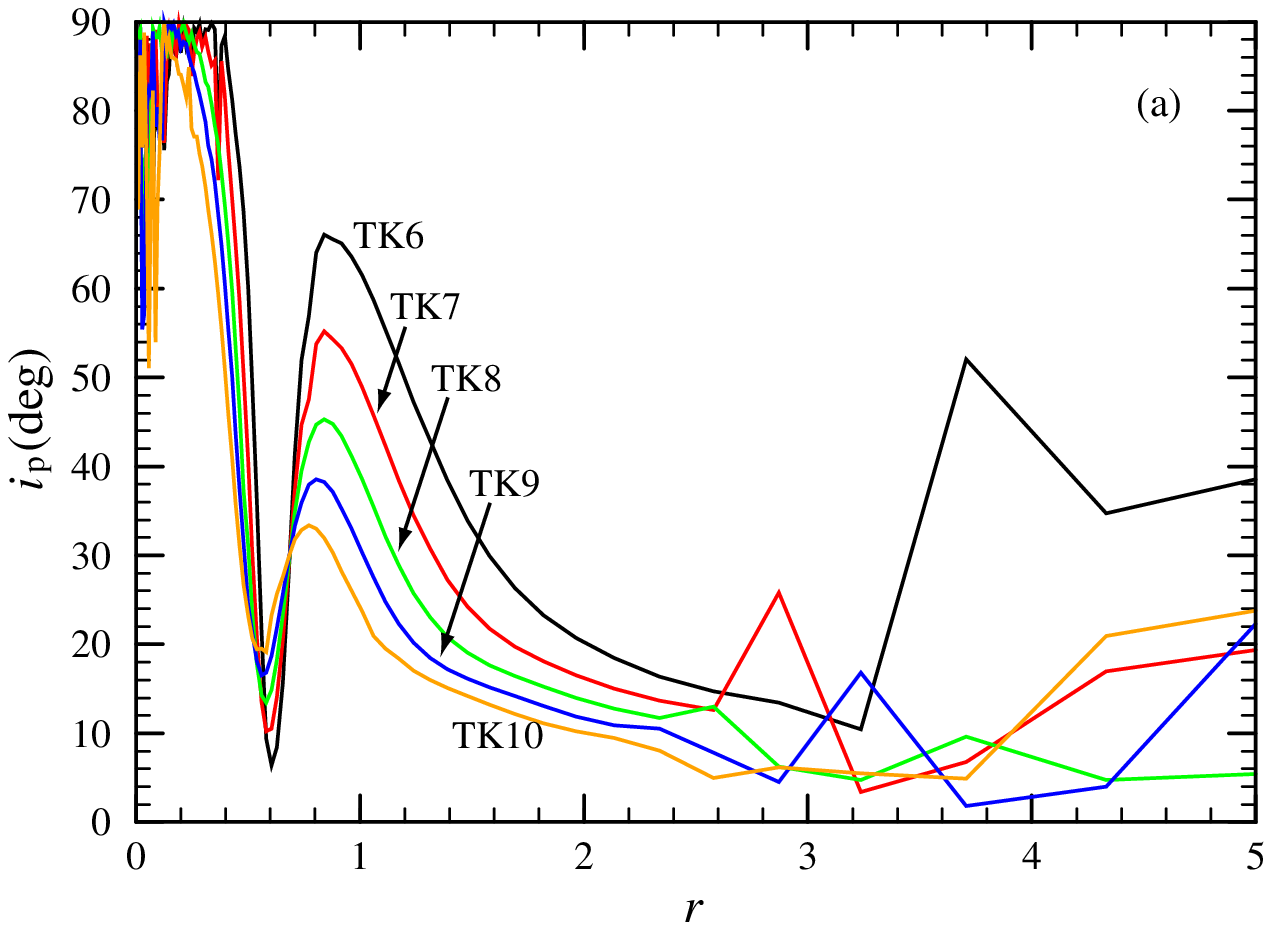}}

\centerline{\includegraphics[width=6.5cm]{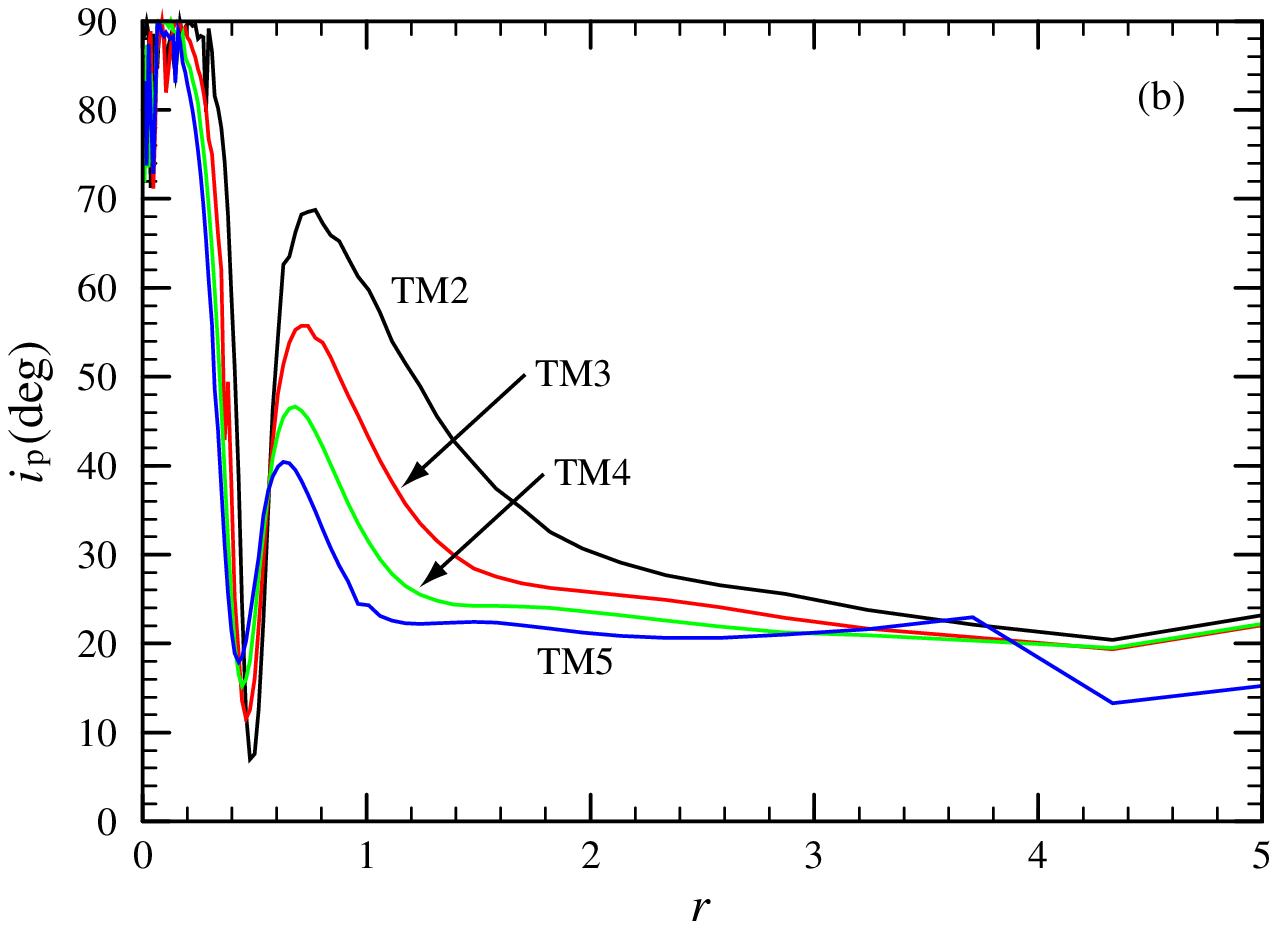}}
\caption{Pitch angles, $i_\mathrm{p}$, measured in degrees
for the most linearly unstable, global two-armed modes
obtained from the linear modal calculations as a function
of radius for (a) the models with Kalnajs's DFs and (b)
those with Miyamoto's DFs.}
\label{fig:pitch_angle}
\end{figure}

\begin{figure}

\centerline{\includegraphics[width=8cm]{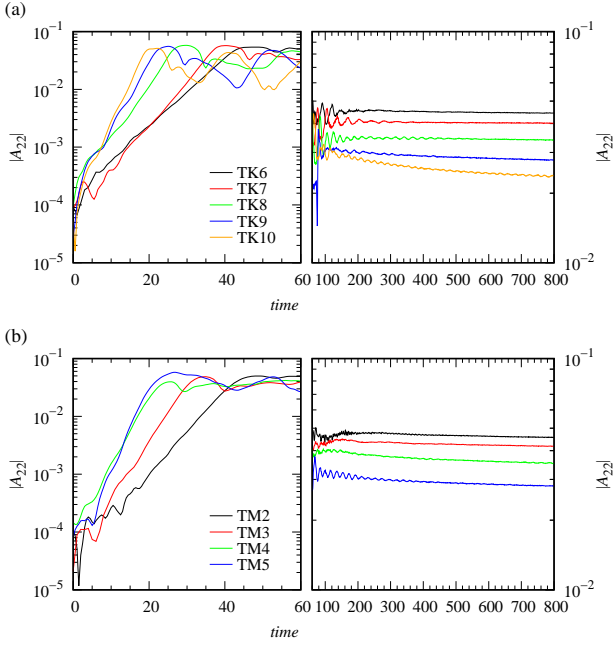}}
\caption{Time evolution of bar amplitude, $|A_\mathrm{22}|$, for
(a) the models with Kalnajs's DFs and (b) those with Miyamoto's
DFs. The labels denote the model names. Note that the scaling
of the abscissa is changed at $t=60$ from the left to the right
panel for each model sequence, and that in accordance with the
change, the ordinate is also re-scaled.}
\label{fig:baramp}
\end{figure}

\begin{figure}
\centerline{\includegraphics[width=7.5cm]{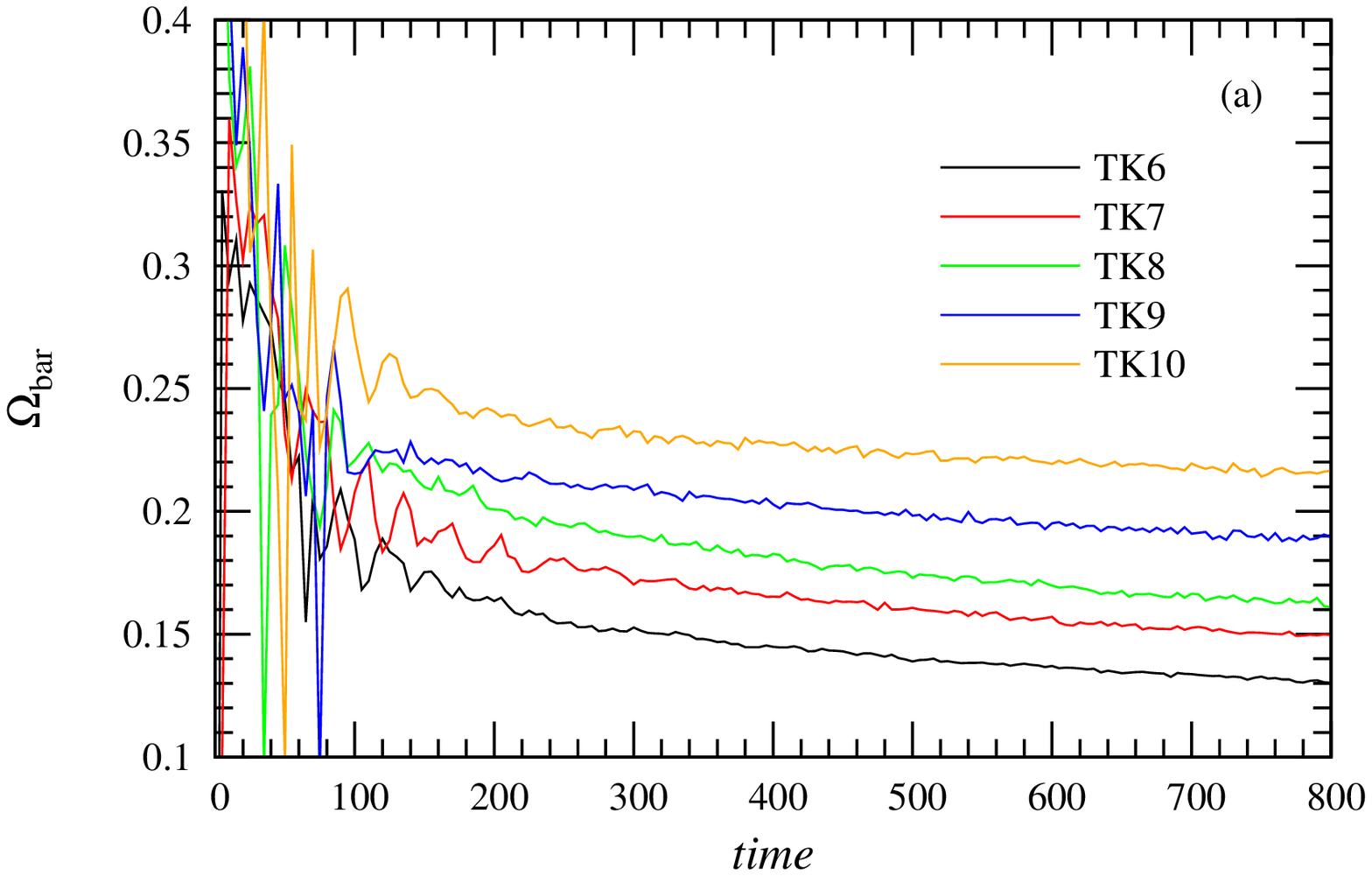}}

\centerline{\includegraphics[width=7.5cm]{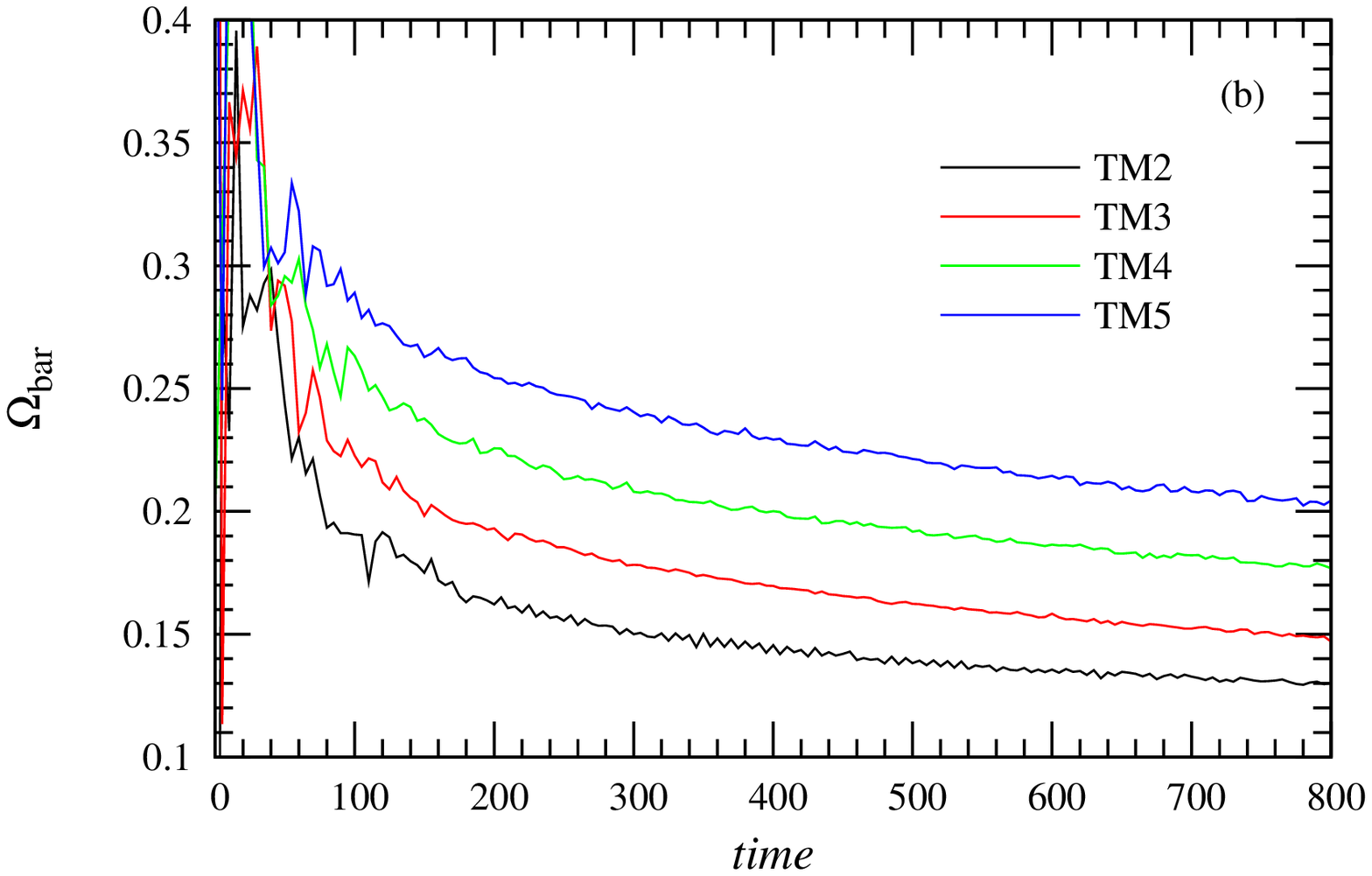}}
\caption{Time evolution of the pattern speed of the two-armed
mode, $\Omega_\mathrm{p}$, which is equivalent to the bar pattern
speed after the bar instability has occurred, for (a) the models
with Kalnajs's DFs and (b) those with Miyamoto's DFs. The labels
denote the model names.}
\label{fig:barpat}
\end{figure}

\begin{figure}
\centerline{\includegraphics[width=7.5cm]{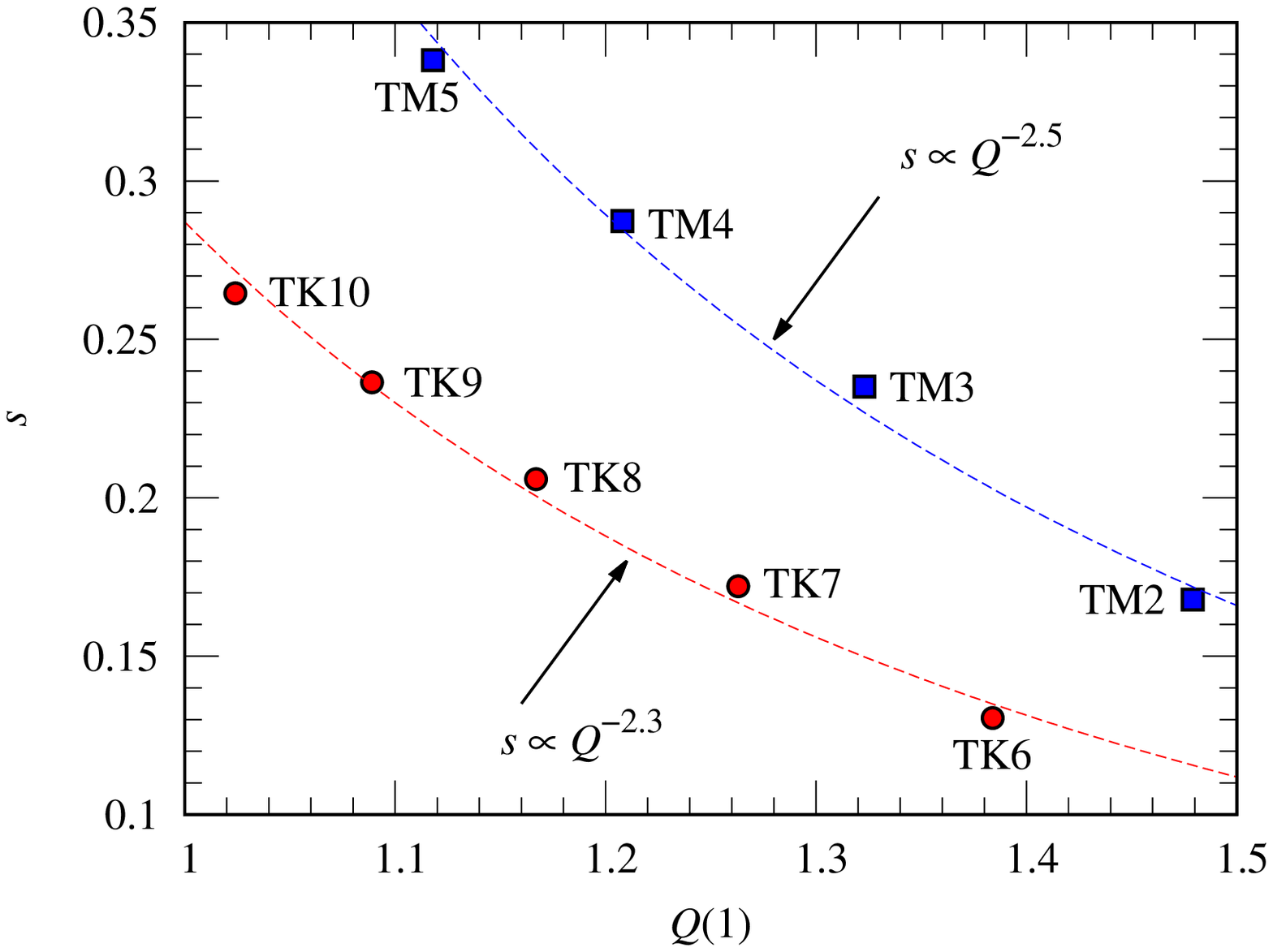}}
\caption{Growth rate, $s$, of the most linearly unstable,
global two-armed mode, obtained from the linear modal
calculations as a function of the value of the Toomre's
$Q$ parameter at the scale length, $Q(1)$. The red dashed
line represents a power-law fit for Kalnajs's DFs while
the blue one denotes that for Miyamoto's DFs.}
\label{fig:q_s}
\end{figure}

\begin{figure}
\centerline{\includegraphics[width=7.5cm]{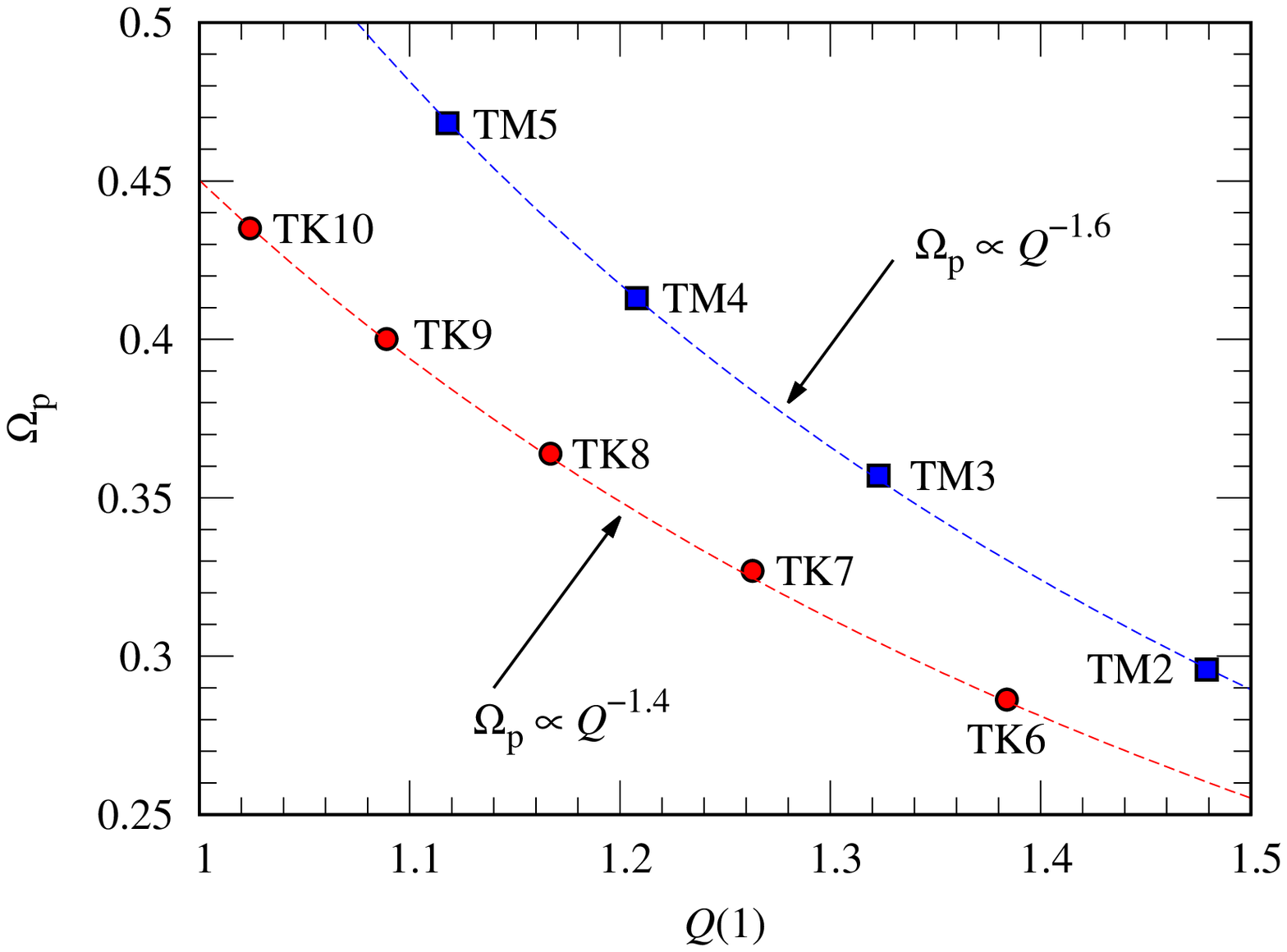}}
\caption{Pattern speed, $\Omega_\mathrm{p}$, of the most
linearly unstable, global two-armed mode, obtained from
the linear modal calculations as a function of the value
of the Toomre's $Q$ parameter at the scale length, $Q(1)$.
The red dashed line represents a power-law fit for Kalnajs's
DFs while the blue one denotes that for Miyamoto's DFs.}
\label{fig:q_omega}
\end{figure}

\begin{figure*}
\centerline{\includegraphics[width=15cm]{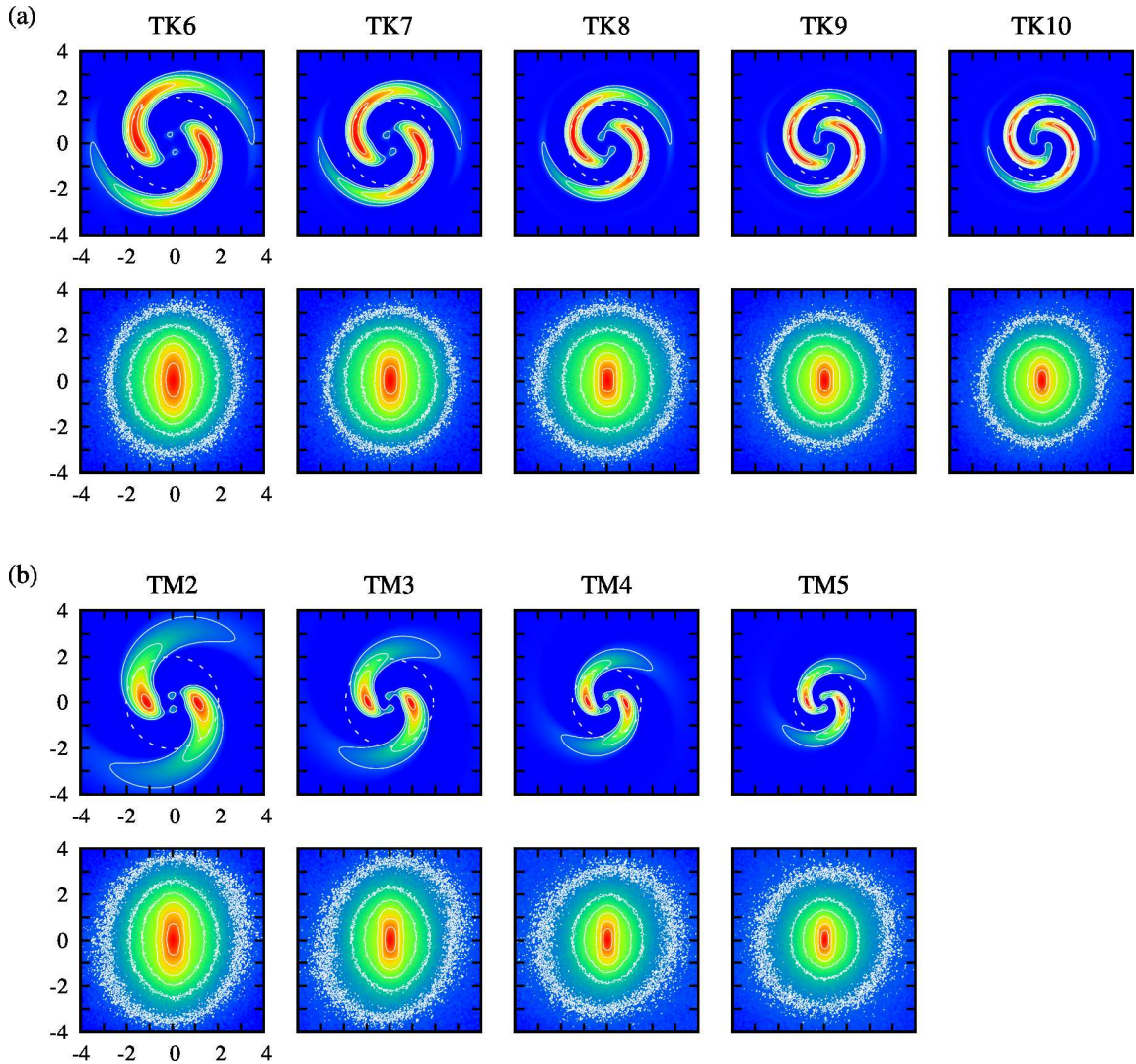}}
\caption{Density contours of the most linearly unstable, global
two-armed modes (upper rows) and those of final bar patterns
(lower rows) for (a) the models with Kalnajs's DFs, and for
(b) those with Miyamoto's DFs. The contours of the most linearly
unstable, global two-armed modes are drawn in white at the 90
per cent, 70 per cent, 50 per cent, $\cdots$, 10 per cent levels
of the peak amplitude on logarithmic scales, while those of the
bar patters are delineated in white at the 90 per cent, 80 per cent,
70 per cent, $\cdots$, 40 per cent levels of the peak amplitude
on logarithmic scales. White dashed circles in the upper rows
show corotation radii. All the spiral and bar patterns rotate
counterclockwise.}
\label{fig:contours}
\end{figure*}

\begin{figure}
\centerline{\includegraphics[width=7.1cm]{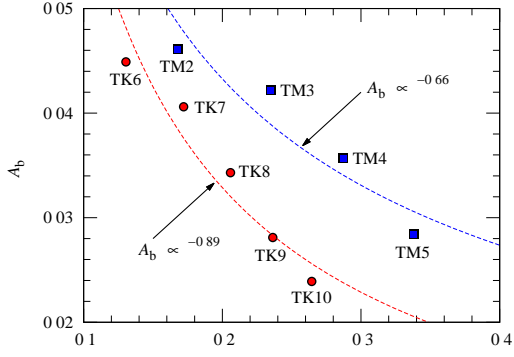}}
\caption{Bar amplitude normalized by the amplitude of the
axisymmetric ring mode averaged over the last 20 time units,
$A_\mathrm{bar}$, obtained from the SCF simulations against
the growth rate of the most linearly unstable, global two-armed mode,
$s$, obtained by solving the linearized collisionless Boltzmann
equation. The red dashed line represents a power-law fit for Kalnajs's 
DFs while the blue one denotes that for Miyamoto's DFs.}
\label{fig:grate_amp}
\end{figure}

\begin{figure}
\centerline{\includegraphics[width=7.1cm]{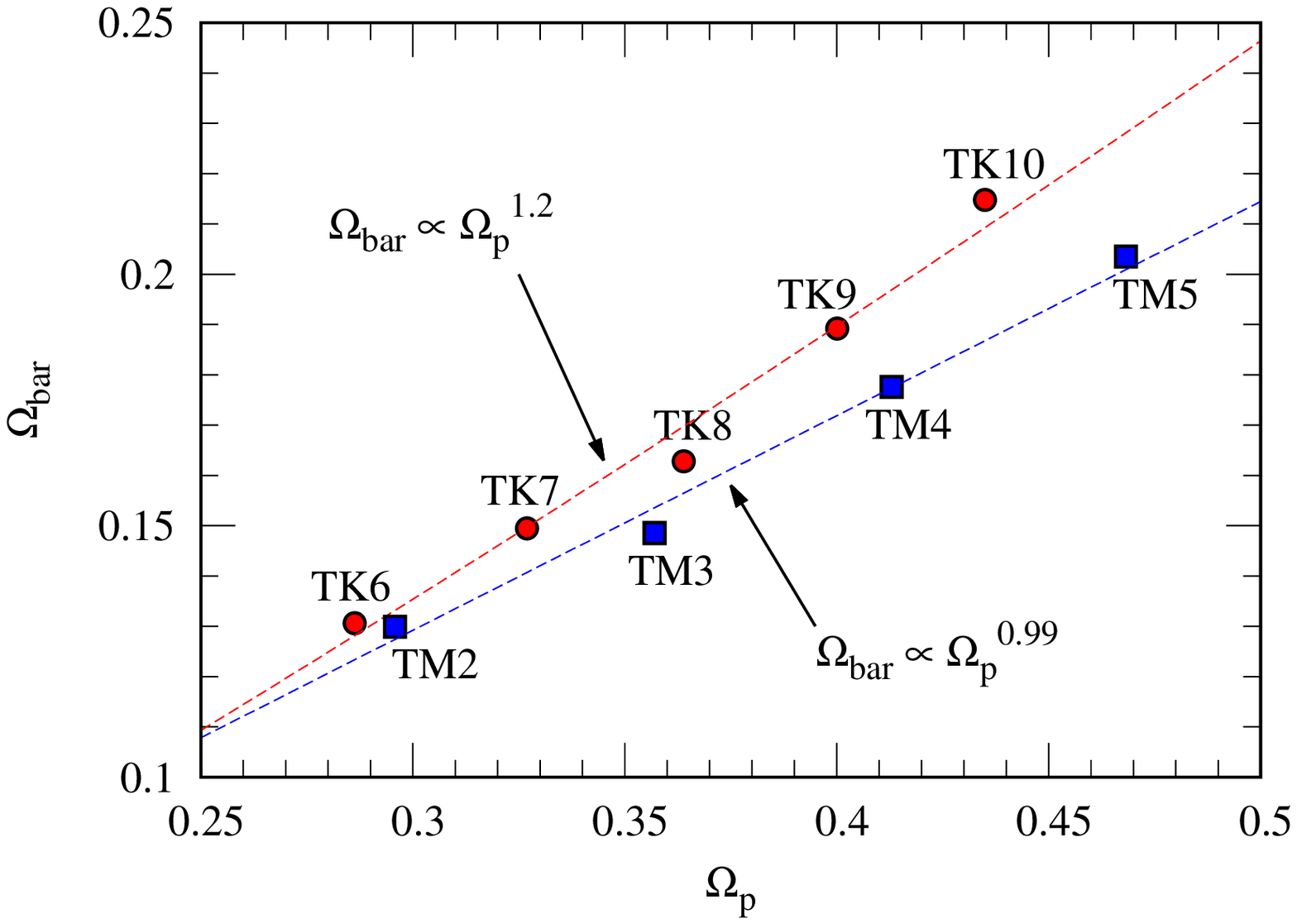}}
\caption{Bar pattern speed, $\Omega_\mathrm{bar}$, obtained from
the SCF simulations against the pattern speed of the most linearly
unstable, global two-armed mode, $\Omega_\mathrm{p}$, obtained by
solving the linearized collisionless Boltzmann equation. The red
dashed line represents a power-law fit for Kalnajs's DFs while
the blue one denotes that for Miyamoto's DFs.}
\label{fig:omega_p}
\end{figure}

We have measured the growth rate, $s$, and pattern speed,
$\Omega_\mathrm{p}$, for each disc model from the linearly
growing stages of the MLUGTAM in the SCF simulations using
a least-squares fitting technique. The results are compared
with those determined from the linear modal calculations based
on the linearized collisionless Boltzmann equation (see equation
\ref{eq:linear_cB}). In Fig.~\ref{fig:eigen_values}, we show
the results obtained with both methods. This figure indicates
that the growth rates and pattern speeds measured from the SCF
simulations are in considerably good agreement with those from
the linear modal calculations. It thus follows that the growing
features in the early stages of the SCF simulations are the
MLUGTAMs inherent in the individual disc models. In addition,
all the $N$-body models were deformed into bars at late times
(see Fig.~\ref{fig:contours}), so that these bars formed in
the SCF simulations can be regarded practically as the end
products of the MLUGTAMs.

We see from Fig.~\ref{fig:eigen_values} that $\Omega_\mathrm{p}$
is almost exactly proportional to $s$
for each type of DF. In a
strict sense, $\Omega_\mathrm{p}\propto s^{0.59}$ for Kalnajs's DFs
while $\Omega_\mathrm{p}\propto s^{0.65}$ for Miyamoto's DFs, if we
fit the data obtained using the linear modal calculations by a
power-law. Furthermore, we find that for a given $s$, Kalnajs's
DFs lead to a higher $\Omega_\mathrm{p}$ than Miyamoto's DFs.
In any case, for the mass profile of the K--T disc, regardless
of the DF employed for a model, a larger $s$ leads to a higher
$\Omega_\mathrm{p}$, and vice versa.

In order to characterize the spiral patterns of the
eigenmodes (see Fig.~\ref{fig:contours}), we calculate
pitch angles along the radius. In Figs~\ref{fig:pitch_angle}(a)
and (b), the pitch angle profiles are presented for the models
with Kalnajs's DFs and those with Miyamoto's DFs, respectively.
These figures show that for each model sequence, the pitch angle
at a given radius within the dominant region of the spiral pattern
decreases as the model parameter, $m_\mathrm{K}$ or $m_\mathrm{M}$,
increases. In addition, the spiral patterns are confined in a
smaller range of radii with increasing model parameter (again
see Fig.~\ref{fig:contours}). Since the growth rate becomes
higher as the model parameter is larger, a more unstable disc
against the two-armed mode results in a more tightly wrapped
and smaller-sized spiral feature for each type of DF.

\subsection{Formation and evolution of bars}\label{subsec:bars}
In Figs~\ref{fig:baramp}(a) and (b), we show
the time evolution of the amplitude of the two-armed (or
the bar) mode, $|A_\mathrm{22}|$, calculated from the SCF
simulations for the models with Kalnajs's DFs and those
with Miyamoto's DFs, respectively. In the early stages
of evolution, a two-armed global mode develops and its
amplitude grows exponentially with time (see the left
panels of Figs~\ref{fig:baramp}a and b). At around the
peak amplitude of each two-armed mode, the bar instability
occurs as a symbolic event of the onset of nonlinear evolution,
so that the excited two-armed spiral mode is deformed into a
bar. In the nonlinear evolution stages, the bar amplitudes remain
nearly constant after they exhibit some fluctuating changes. On
the other hand, after the bars have fully grown up, their pattern
speeds decrease over time as presented in Figs~\ref{fig:barpat}(a)
and (b) for the models with Kalnajs's DFs
and those with Miyamoto's DFs, respectively, because
of the angular momentum exchange between the bar pattern
and disc stars \citep{little91}.

\subsection{Correlations}\label{subsec:correlations}
\begin{figure}
\centerline{\includegraphics[width=7.1cm]{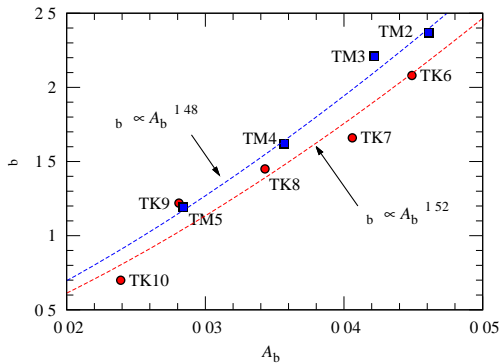}}
\caption{Final bar half-size, $L_\mathrm{bar}$, as a function
of the normalized bar amplitude, $A_\mathrm{bar}$. Here,
the bar amplitude is obtained from an average over the
last bar rotation period, and normalized by the amplitude
of the axisymmetric ring mode averaged over the same period,
while the bar half-size is calculated from a combination of
the Fourier components of the disc surface density (see the text).
The red dashed line represents a power-law fit for Kalnajs's 
DFs while the blue one denotes that for Miyamoto's DFs.}
\label{fig:barsize}
\end{figure}

\begin{figure}
\centerline{\includegraphics[width=7.1cm]{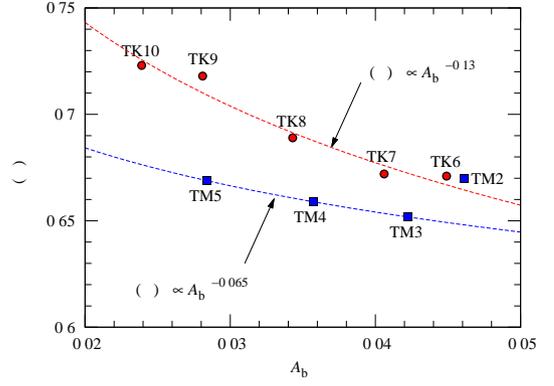}}
\caption{Final smallest axis ratio, $(b/a)_\mathrm{min}$,
of the bar along the bar major axis as a function of
the normalized bar amplitude, $A_\mathrm{bar}$. The bar
amplitude is obtained as described in Fig.~\ref{fig:barsize}.
The red dashed line represents a power-law fit for Kalnajs's
DFs while the blue one denotes that for Miyamoto's DFs
without model TM2.}
\label{fig:bar_axis}
\end{figure}

\begin{figure*}
\centerline{\includegraphics[width=16cm]{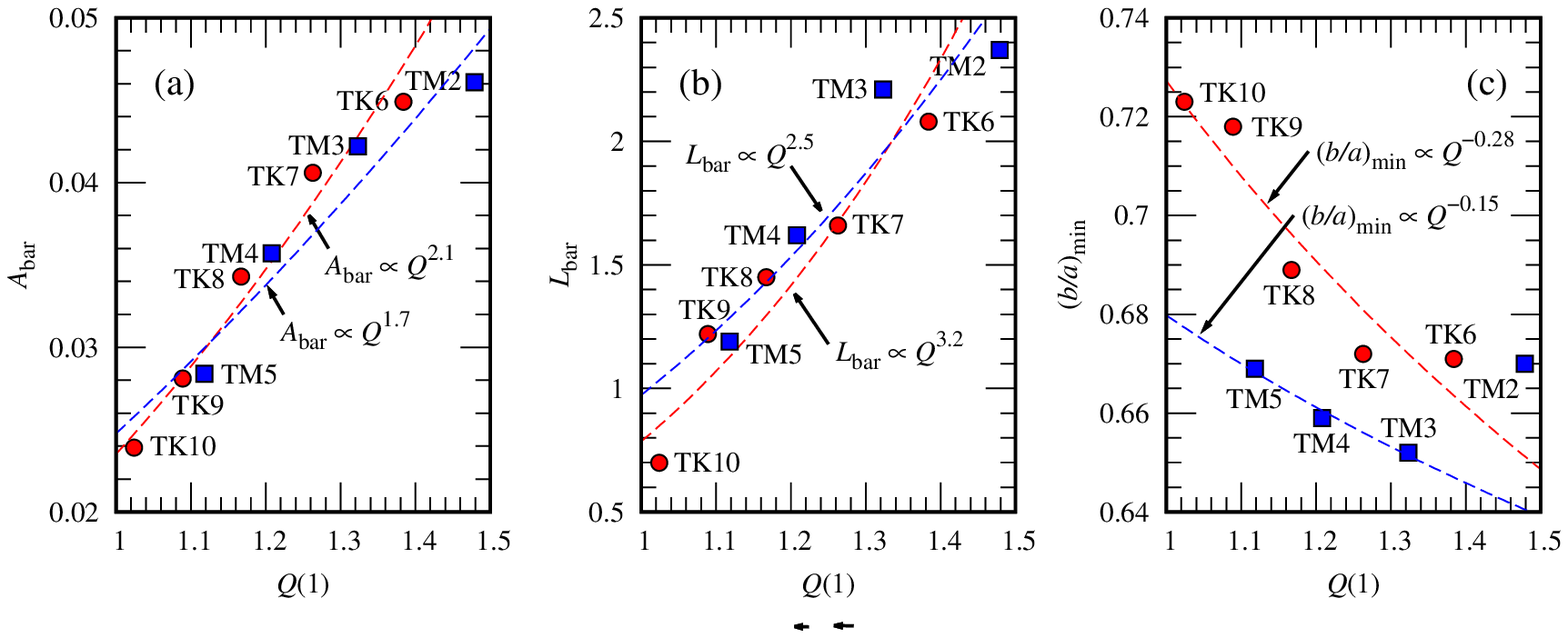}}
\caption{Resulting normalized bar amplitude, $A_\mathrm{bar}$, (a)
bar half-size, $L_\mathrm{bar}$, (b) and minimum bar axis ratio,
$(b/a)_\mathrm{min}$, (c) as a function of the Toomre's $Q$ parameter
at the scale length, $Q(1)$. In each panel, the circles show
the models with Kalnajs's distribution functions, while the
squares denote those with Miyamoto's DFs.
The red dashed lines represent power-law fits for Kalnajs's
DFs while the blue ones denote those for Miyamoto's DFs,
where model TM2 is excluded when a power-law fit is applied
to $(b/a)_\mathrm{min}$.}
\label{fig:bar_properties}
\end{figure*}

In Figs~\ref{fig:q_s} and \ref{fig:q_omega}, we show, respectively,
the growth rate, $s$, and pattern speed, $\Omega_\mathrm{p}$, of
the MLUGTAM as a function of the Toomre's $Q$ value at the scale
length, $Q(1)$. Power-law fits indicate that $s\propto Q^{-2.3}$
and $\Omega_\mathrm{p} \propto Q^{-1.4}$ for Kalnajs's DFs while
$s\propto Q^{-2.5}$ and $\Omega_\mathrm{p}\propto Q^{-1.6}$ for
Miyamoto's DFs. We find from these figures that in the linear
regime, both $s$ and $\Omega_\mathrm{p}$ of the MLUGTAM decrease
as $Q(1)$ increases. We also note that for a given $Q(1)$,
Kalnajs's DFs show smaller $s$ and $\Omega_\mathrm{p}$ than
Miyamoto's DFs. Thus, for each model sequence, the
eigenvalue representing the growth rate and pattern speed
of the MLUGTAM is specified by $Q(1)$. In order to examine the
correspondence between the linear and nonlinear regimes,
the final density contours of the bars are presented in
Fig.~\ref{fig:contours}, together with those of the corresponding
MLUGTAMs. By comparing these paired density contours along each
model sequence, it appears that the properties of the bars reflect
those of the spiral patterns of the MLUGTAMs. Since the MLUGTAM
is identified with the growth rate and pattern speed, both of which
are correlated with $Q(1)$ as exhibited in Figs~\ref{fig:q_s} and
\ref{fig:q_omega}, the correspondence between the properties of
the bars and those of the spiral density patterns illustrated in
Fig.~\ref{fig:contours} suggests that the bars can be characterized
by the typical Toomre's $Q$ value. In what follows, we will reveal
correlations in the quantities that prescribe the bar, finally
demonstrating how these quantities are correlated with $Q(1)$.

In Fig.~\ref{fig:grate_amp}, we show a correlation between
the growth rate obtained from the linear modal calculations
and the bar amplitude, $A_\mathrm{bar}$, from the SCF simulations.
In plotting this figure, we first define the bar pattern speed,
$\Omega_\mathrm{bar}$, as an average over that period of the
last 20 time units (from $t=780$ to $t=800$) which corresponds
to roughly half the bar rotation period of each model. Next,
we calculate the bar amplitude, $\overline{|A_\mathrm{22}|}$,
as an average over the bar rotation period derived from
$\Omega_\mathrm{bar}$ and then, normalize it by the amplitude
of the axisymmetric ring mode, $\overline{|A_\mathrm{00}|}$,
again averaged for that same period. Hereafter, we refer to
the bar amplitude as that calculated this way, unless otherwise
mentioned, that is, $A_\mathrm{bar}=\overline{|A_\mathrm{22}|}
/\overline{|A_\mathrm{00}|}$. Fig.~\ref{fig:grate_amp} indicates
that for each type of DF, the bar amplitude decreases almost
linearly with increasing growth rate. More precisely, a power-law
fit leads to $A_\mathrm{bar}\propto s^{-0.89}$ for Kalnajs's
DFs while it results in $A_\mathrm{bar}\propto s^{-0.66}$ for
Miyamoto's DFs. We see from Fig.~\ref{fig:grate_amp} that for
a given growth rate, Kalnajs's DFs produce a lower amplitude
bar than Miyamoto's DFs.

In Fig.~\ref{fig:omega_p}, we present a correlation between
$\Omega_\mathrm{p}$ of the MLUGTAM obtained from the linear modal
calculations, and $\Omega_\mathrm{bar}$ from the SCF simulations.
We find from this figure that $\Omega_\mathrm{bar}$ is almost
proportional to $\Omega_\mathrm{p}$. In reality, power-law fits
show that $\Omega_\mathrm{bar}\propto {\Omega_\mathrm{p}}^{1.2}$
for Kalnajs's DFs while $\Omega_\mathrm{bar}\propto
{\Omega_\mathrm{p}}^{0.99}$ for Miyamoto's DFs. In addition,
for a given $\Omega_\mathrm{p}$, Kalnajs's DFs generate a higher
$\Omega_\mathrm{bar}$ than Miyamoto's DFs.

Figs~\ref{fig:grate_amp} and \ref{fig:omega_p} imply that
$A_\mathrm{bar}$ and $\Omega_\mathrm{bar}$ are closely connected
to the eigenvalue of the disc model, from which $s$ and
$\Omega_\mathrm{p}$ are calculated. Consequently, from a
quantitative point of view, the properties of a bar
reflect those of its corresponding MLUGTAM.

As a bar property, we first measure the bar length,
$L_\mathrm{bar}$. Following \citet{ohw90}, we calculate,
as a bar half-size, the radius of the bar region in
which the value of $(I_0+I_2+I_4+I_6)/(I_0-I_2+I_4-I_6)$
exceeds 2.0, where $I_m$ ($m=0,\, 2,\, 4,$ and 6)
is the $m$th Fourier component
of the disc surface density. We regard this radius as
$L_\mathrm{bar}$. In Fig.~\ref{fig:barsize}, $L_\mathrm{bar}$
at $t=800$ is plotted against $A_\mathrm{bar}$. From this
figure, we find that $L_\mathrm{bar}$ is roughly proportional
to $A_\mathrm{bar}$, irrespective of the type of DF. As a
precise description made by power-law fits, $L_\mathrm{bar}
\propto {A_\mathrm{bar}}^{1.52}$ for Kalnajs's DFs while
$L_\mathrm{bar}\propto {A_\mathrm{bar}}^{1.48}$ for Miyamoto's
DFs. It thus turns out that a lower amplitude bar is shorter
in length.

As another bar property, we measure the roundness of the bar,
which is defined as the shortest axis ratio along the bar major
axis, $(b/a)_\mathrm{min}$. In doing so, the axis ratios of a
bar along the radius are determined by calculating the principal
moment of inertia tensor for particles included within a specified
radius, and then, the value of this moment is assigned to the
axis ratio at that radius. Fig.~\ref{fig:bar_axis} shows the
roundness of the bar as a function of $A_\mathrm{bar}$. This
figure indicates that $(b/a)_\mathrm{min}$ is roughly linearly
decreasing as $A_\mathrm{bar}$ increases except for that model
TM2 in the sequence of Miyamoto's DFs which is the hottest model
characterized by the largest typical $Q$ value. A power-law fit
tells us that $(b/a)_\mathrm{min}\propto {A_\mathrm{bar}}^{-0.13}$
for Kalnajs's DFs while $(b/a)_\mathrm{min}\propto
{A_\mathrm{bar}}^{-0.065}$ for Miyamoto's DFs without
model TM2. Unlike $L_\mathrm{bar}$, $(b/a)_\mathrm{min}$
depends on the functional form of DF in the sense that
Miyamoto's DFs lead to a more elongated bar than Kalnajs's
DFs for a given $A_\mathrm{bar}$. At any rate, for each
type of DF, a lower amplitude bar is rounder in shape
except for model TM2 whose bar shows the highest amplitude
in the sequence of Miyamoto's DFs.

As exhibited in Fig.~\ref{fig:grate_amp}, $A_\mathrm{bar}$
is correlated with $s$, while $L_\mathrm{bar}$ and
$(b/a)_\mathrm{min}$ are correlated with $A_\mathrm{bar}$,
which is shown in Figs~\ref{fig:barsize} and \ref{fig:bar_axis},
respectively. Consequently, it follows that $L_\mathrm{bar}$
and $(b/a)_\mathrm{min}$ are, respectively, also correlated
with $s$. On the other hand, as revealed in Fig.~\ref{fig:q_s},
$s$ is well-correlated with $Q(1)$, and so, we can infer that
$A_\mathrm{bar}$, $L_\mathrm{bar}$ and $(b/a)_\mathrm{min}$ are,
respectively, specified by $Q(1)$. In fact, as demonstrated in
Fig.~\ref{fig:bar_properties}, in which $A_\mathrm{bar}$,
$L_\mathrm{bar}$, and $(b/a)_\mathrm{min}$ are plotted as
a function of $Q(1)$, we see that $A_\mathrm{bar}$ and
$L_\mathrm{bar}$ are well-correlated with $Q(1)$, almost
regardless of the functional form of DF, while $(b/a)_\mathrm{min}$
is roughly correlated with $Q(1)$ and depends on the DF used.
With the help of a power-law fit, we find that $A_\mathrm{bar}
\propto Q^{2.1}, L_\mathrm{bar}\propto Q^{3.2}$, and
$(b/a)_\mathrm{min}\propto Q^{-0.28}$ for Kalnajs's DFs
while $A_\mathrm{bar}\propto Q^{1.7}, L_\mathrm{bar}\propto Q^{2.5}$,
and $(b/a)_\mathrm{min}\propto Q^{-0.15}$ for Miyamoto's DFs, where
model TM2 is excluded for fitting $(b/a)_\mathrm{min}$.
We note again that for $(b/a)_\mathrm{min}$, model TM2 deviates
from the correlation with $Q(1)$.

As an additional remark, no tight correlation
has been found between the final $Q$ and the bar properties.

\section{Discussion}\label{sec:discussion}
We have found that as a disc evolves, the MLUGTAM
is growing in the linear phases, and that it eventually
turns into a bar via the bar instability. We can
thus infer that the properties of a bar such as the amplitude,
length, and axis ratio are related to those of the MLUGTAM.
In fact, as presented in Figs~\ref{fig:grate_amp} and
\ref{fig:omega_p}, the amplitude and pattern speed of
a bar are, respectively, well-correlated with the growth
rate and pattern speed of the MLUGTAM. Therefore, we
discuss below the relation between the MLUGTAM and the bar
properties, and thereby we will try to unravel the origin
of the correlation, revealed in Fig.~\ref{fig:bar_properties},
between the initial $Q(1)$ and the bar properties.

Fig.~\ref{fig:contours} demonstrates that the MLUGTAM is
confined to a smaller radius as the initial $Q(1)$ decreases.
A more confined two-armed pattern could be considered to
be formidable to provide gravitational influences farther away,
unless its amplitude is sufficiently large. Consequently,
a bar which is produced by a more confined MLUGTAM would result
in a lower amplitude state owing to the difficulty in attracting many
more masses in the bar, leading to the correlation that the
bar amplitude decreases with decreasing $Q(1)$, as exhibited
in Fig.~\ref{fig:bar_properties}a.

Regarding the bar length, $L_\mathrm{bar}$, bar-supporting
orbits do not exist beyond the bar corotation radius
\citep{contopoulos80}, $r_\mathrm{CR}$, at which the
bar pattern speed, $\Omega_\mathrm{bar}$, is equal to
the angular speed of a star on a circular orbit, $\Omega$.
Consequently, assuming that the orbital
content is similar in all cases, we can use $r_\mathrm{CR}$ as a
measure of $L_\mathrm{bar}$. In fact, from observations,
\citet{cuomo20} have found the correlation that longer
bars have larger corotation radii. Since $\Omega$ is
a decreasing function of the radius, $r_\mathrm{CR}$
becomes smaller with increasing $\Omega_\mathrm{bar}$.
\citet{cuomo20} have found this relation in real barred
galaxies as well. In addition, as presented in
Fig.~\ref{fig:omega_p}, $\Omega_\mathrm{bar}$ is approximately
proportional to the pattern speed of the MLUGTAM, $\Omega_\mathrm{p}$.
Accordingly, as $\Omega_\mathrm{p}$ is smaller, $\Omega_\mathrm{bar}$
is smaller, so that $r_\mathrm{CR}$, as a result $L_\mathrm{bar}$ also,
is larger as $\Omega_\mathrm{p}$ is smaller. On the other hand,
Fig.~\ref{fig:q_omega} indicates that $\Omega_\mathrm{p}$ decreases
as the initial $Q(1)$ increases. In this way, $L_\mathrm{bar}$
increases with increasing $Q(1)$, as shown in
Fig.~\ref{fig:bar_properties}b.

Fig.~\ref{fig:contours} demonstrates also that a more rounder
bar is produced by a more tightly wrapped MLUGTAM. From the
density wave theory, the $m$-armed wave satisfies the dispersion
relation \citep[e.g.][]{bt08} represented by
\begin{eqnarray}
\label{eq:dispersion}
(\omega - m\Omega_\mathrm{p})^2 & = & c^2\,k^2-2\uppi G\mu\,|k| + \kappa^2\\
& = & c^2\left(|k|-\frac{\uppi G\mu}{c^2}\right)^2 +
      \kappa^2\left(1-\frac{1}{Q^2}\right),\nonumber
\end{eqnarray}
where $\omega$ is the angular frequency, $k$ is the radial
wavenumber, $c$ is the radial velocity dispersion, and the
Toomre's $Q$ is defined with a fluid approximation as
\begin{equation}
Q = \frac{\kappa c}{\uppi G\mu}.
\end{equation}
If the disc is unstable, equation (\ref{eq:dispersion}) suggests
that the most unstable wavelength, $\lambda_\mathrm{u}$, corresponding
to the most unstable wavenumber, $|k_\mathrm{u}|=\uppi G\mu/c^2$, is
provided by
\begin{equation}
\lambda_\mathrm{u} = \frac{2\uppi}{|k_\mathrm{u}|} = \frac{2c^2}{G\mu}
          = \frac{2\uppi^2 G\mu}{\kappa^2}Q^2.
\end{equation}
Since the mass profiles used are those of the K--T discs,
$\mu$ and $\kappa$ do not change from model to model. It
thus follows that $\lambda_\mathrm{u}$ is shorter with decreasing
$Q$. The shorter radial wavelength means a more tightly wrapped
spiral. Therefore, the pitch angle becomes smaller as $Q$
decreases. Indeed, from Fig.~\ref{fig:pitch_angle}, we find
that the pitch angle decreases as $Q(1)$ decreases for each
model sequence. The deviation of the force field from the
axisymmetry is considered to be smaller for a tightly wrapped
spiral than for a loosely wrapped one. Thus, it is conceivable
that a more tightly wrapped spiral could result in a less
violent change in the force field along the azimuthal direction
when the bar instability occurs, and so, the produced bar could
be rounder as the MLUGTAM is more tightly wrapped. Consequently,
the correlation between the axis ratio and the
initial $Q(1)$ might emerge as revealed in
Fig.~\ref{fig:bar_properties}c.


We now know from observations that the bar length decreases
from early- to late-type barred galaxies \citep{erwin05} while
longer bars have larger amplitudes \citep{eekb07, diaz16a, guo19,
cuomo19, cuomo20}. It thus follows that the bar amplitude
decreases from SBa to SBc or SBd. Furthermore, observations
show that larger amplitude bars are more morphologically elongated
\citep{menendez07, hoyle11}, which means that the minor-to-major
axis ratio of a bar becomes larger from SBa to SBc or SBd.
Combining these consequences with the correlations exhibited
in Figs~\ref{fig:bar_properties}(b) and (c), we can infer
that the Hubble sequence for barred galaxies could be the
sequence of decreasing $Q$ from SBa to SBc or SBd.

On the other hand, if we rely on the
observations that the fraction of barred galaxies increases
with time \citep{sheth08, melvin14}, as described in
\mbox{Section \ref{sec:introduction}}, the spiral structure
in non-barred galaxies that we observe at present might be
the MLUGTAM still on the stage of the linear growth, which
would be deformed into barred galaxies sometime in the future.
Indeed, \citet{bertin89} demonstrated that unstable global
modes can at least generate the spiral morphology of all
Hubble types, irrespective of non-barred or barred spirals.
Actually, large-scale smooth symmetric arms which are
reminiscent of a globally unstable mode are detected
in the disc for the old stellar population at near-infrared
wavelengths \citep{block94}. In addition, as shown by
observations, if the difference
in the properties between non-barred and
barred galaxies is basically whether a bar exists
or not, the Hubble sequence for non-barred galaxies
might also represent the sequence of decreasing $Q$
from Sa to Sc or Sd. If this is the
case, the $Q$ value at some radius like the disc scale
length might give a clue to the initial typical $Q$ for
barred galaxies under the assumption that the spiral structure
in non-barred galaxies is indicative of the MLUGTAM before
transforming itself into a bar. In this sense, it will be
important to obtain $Q$ values observationally for non-barred
galaxies along the Hubble sequence. However, our results
lead to the consequence that the pitch angle of the MLUGTAM
decreases with decreasing $Q$ (see Figs~\ref{fig:pitch_angle}
and \ref{fig:contours}), which indicates that if the Hubble
sequence is the sequence of decreasing $Q$, the spiral
arm is more tightly wrapped from Sa to Sc or Sd, contrary
to the real Hubble sequence. This discrepancy suggests
that the MLUGTAM itself might not correspond directly
to the observed spiral pattern. In fact,
recent numerical simulations have revealed that spiral
arms are not global steady patterns like the MLUGTAMs
studied here but transient features \citep{fujii11,
wada11, baba13} repeatedly excited by swing amplification
\citep{toomre81}. Therefore, considering
that modal calculations expose the existence of numerous
unstable modes in self-gravitating discs \citep[e.g.][]{av83},
we will need to make clear the relation between the spiral
structure and the MLUGTAM in order to confirm whether
the Hubble sequence is the sequence of decreasing $Q$.
Even though the initial typical $Q$ is the key ingredient
only to the Hubble sequence for barred galaxies, it will
be significant to verify our findings using those realistic
disc models with finite thickness which are embedded in
live dark matter haloes.

Our results are obtained from extremely ideal disc models
which are different from the galactic discs in the real
Universe. For example, we have adopted K--T discs as
a mass profile because the exact equilibrium DFs are
known, in spite of the fact that real disc galaxies
are represented by exponential surface density profiles
\citep{freeman70}. In particular, each of our disc models
lacks a dark matter halo which is assumed to surround a
disc, so that the effects of wave-particle interactions
between a bar mode and halo particles \citep{lia02} are
neglected in the present study. Such resonant interactions
can amplify the bar strength, especially for a massive,
centrally concentrated halo \citep{lia02}. In addition,
the motions of stars in our simulations are restricted
to a single plane. As a result, the effects of buckling
instabilities intrinsic in three-dimensional discs
\citep{raha91, victor04, martinez04} are not taken
into account, although such instabilities reduce the
bar strength and can dissolve a bar in non-violent
buckling cases \citep{collier20}. Thus, the initial
typical $Q$ would not be the only factor that determines
the bar amplitude. Accordingly, it is likely that the
correlation between the bar amplitude and the initial
typical $Q$ will be altered from that shown in
Fig.~\ref{fig:bar_properties}a for realistic
disc galaxy models. Similarly, in addressing real
barred galaxies, a certain modification might be
required for our findings that the bar properties
such as the length and axis ratio are closely connected
to the initial typical $Q$, which are represented in
Figs~\ref{fig:bar_properties}(b) and (c). However,
the correlations, which we have revealed here, between
$L_\mathrm{bar}$ and the bar amplitude, $A_\mathrm{bar}$,
depicted in Fig.~\ref{fig:barsize} and between the
axis ratio, $(b/a)_\mathrm{min}$, and $A_\mathrm{bar}$,
illustrated in Fig.~\ref{fig:bar_axis} are consistent
with those observed in real barred galaxies
\citep{erwin05, eekb07, menendez07, hoyle11}.
Therefore, the correlations shown in Figs
\ref{fig:bar_properties}(b) and (c) might
hold for real barred galaxies.

\section{Conclusions}\label{sec:conclusions}
We have obtained the MLUGTAMs of razor-thin K--T disc models
constructed with the exact equilibrium DFs by solving the
linearized collisionless Boltzmann equation as an initial
value problem. In addition, we have carried out $N$-body
simulations with a softening-free SCF code using the same
disc models. Putting both results together, we have identified
that the growing feature in the early phases of disc evolution,
which is finally disfigured to form a bar through the bar
instability, is the MLUGTAM.

From the SCF simulations, we have confirmed that the resulting
bars show the correlations observed in real barred galaxies such
that the length increases and the axis ratio decreases as the
amplitude increases. By demonstrating that the amplitude and
pattern speed of a bar are, respectively, well-correlated with
the growth rate and pattern speed of the MLUGTAM, we have shown
that the correlations found in the simulated bars root in the
eigenvalue of that MLUGTAM. The properties of the bar formed
by the bar instability thus reflect those of its corresponding
MLUGTAM. Furthermore, we have also shown that the growth rate
and pattern speed of the MLUGTAM are well-correlated with
the initial $Q$ value at the scale length of the disc.
Consequently, we have revealed that the amplitude and the
length increases while the axis ratio in itself decreases,
as that $Q$ value increases. Therefore, we conclude that
this typical $Q$ value is the determinant of the bar properties.
On the basis of this finding, we suggest that the Hubble sequence
for barred galaxies, SB, might be the sequence of decreasing $Q$.

\section*{Acknowledgements}
The author would like to thank Takao Fujiwara for providing
him with the linearized collisionless Boltzmann code and
Roland \mbox{Jesseit} and Andreas Burkert for fruitful discussions.
He is indebted to Panos Patsis for his careful reading of
the manuscript and constructive comments on it. He acknowledges
Lars Hernquist for the encouragement of this research and Keigo
Nitadori for optimizing the two-dimensional SCF code used in
this paper for parallelization. Thanks are also due to Shoji
Kato for a valuable suggestion on the understanding of the
relation between the size and amplitude of a bar.
The author would also like to thank the anonymous referee for
the valuable comments and constructive suggestions that have
helped to improve the manuscript. Numerical computations were
carried out on Cray XT4 at the Centre for Computational Astrophysics
(CfCA), the National Astronomical Observatory of Japan.

\section*{Data Availability}
The source code for solving the linearized collisionless Boltzmann
equation and the SCF code used for the simulations as well as the
models and the simulation data underlying this article will be shared
on reasonable request to the corresponding author.


\bibliographystyle{mnras}
\bibliography{hozumi} 


\bsp	
\label{lastpage}
\end{document}